\documentclass[prl,superscriptaddress,showpacs,floatfix,twocolumn,amsmath,amsfonts,amssymb,10pt]{revtex4}
\usepackage{graphics}
\usepackage{epsfig}
\usepackage{amssymb}
\usepackage{color}
\begin{document}
\date{\today}

\title{Discrete solitons of spin-orbit coupled Bose-Einstein condensates in optical lattices  }

\author{Mario Salerno}
\affiliation{ Dipartimento di Fisica ``E.R. Caianiello'', CNISM and INFN
- Gruppo Collegato di Salerno, Universit\'a di Salerno, Via Giovanni
Paolo II, 84084 Fisciano (SA), Italy}
\author{Fatkhulla Kh. Abdullaev}
\affiliation{Deparment of Physics, Kulliyyah of Science, International
Islamic \\University of Malaysia, 25200 Kuantan, Pahang, Malaysia}
\begin{abstract}
We study localized nonlinear excitations of a dilute Bose-Einstein condensate (BEC) with spin-orbit coupling in a deep optical lattice (OL).
We use Wannier functions to derive a tight-binding model that includes the spin-orbit coupling (SOC) at the discrete level in the form of a generalized discrete nonlinear Sch\"odinger equation. Spectral properties are investigated and the existence and stability of discrete solitons and breathers  with different symmetry properties with respect to the OL is demonstrated. We show that the symmetry of the modes can be changed from on-site to inter-site and to asymmetric modes simply by changing  the interspecies interaction. Asymmetric modes appear to be novel modes intrinsic of the SOC.
\end{abstract}
\maketitle

{\it Introduction.} Presently there is a growing interest in the study of Bose-Einstein condensates (BEC) in the presence of non abelian  gauge fields that mimic magnetic interactions. In particular, spin-orbit couplings (SOC) of Rashba and Dresselhaus types and their combinations, have been recently realized in binary BEC mixtures in the presence of trapping potentials $V(x)$ of different types\cite{LJS}(see the review articles\cite{GS,rev2}). As  well known, the SOC represents a major source of intra-atomic  magnetic interaction. In solid state physics it plays an important role in  the magnetism of solids, well described in terms of individual  ions, as it is for earth rare insulators. In general, however, in solids the SOC is a rather weak source of magnetic interaction (largely superseded by the electrostatic effects, which is  impossible to manage/enhance.

The situation is quite different in Bose-Einstein condensates where  a variety of forms of synthetic spin-orbit couplings can be easily generated  by external laser fields and  the strength can be  easily  controlled. BEC with SOC, indeed,  now represent ideal systems to explore interesting phenomena that are difficult to achieve in  solid state,  such as  new quantum phases with unusual magnetic properties,  existence of stripe modes\cite{stripe}, fractional topological insulators\cite{phase,phase1,phase2}, Majorana fermions, etc.

The interplay of the spin-orbit coupling with the  interatomic interactions (nonlinearity) and the periodicity of the OL also leads to the existence of  SOC solitons\cite{sol2,sol3}. Gap solitons in BEC with periodic Zeeman field were recently discussed  in \cite{KKA-PRL13} (see also the review of BEC in OL with SOC in  \cite{kit0}). In particular, different settings for the spatial periodicity have been considered: periodicity in each separate component\cite{Sakaguchi,Zhang}, periodicity in the Raman coupling\cite{DKP}, periodicity in the Zeeman field\cite{KKA-PRL13,EOUFTO,JLWBPS,ZMZ}. Dispersion relations of one-dimensional BEC with SOC in OL were experimentally investigated in \cite{exp1} and existence of flat bands and superfluidity in BEC with SOC  demonstrated in \cite{Zhang}. Array of vortex lattices in shallow optical lattices  \cite{Sakaguchi} and vortices in 2D optical lattices \cite{LKK} were also investigated. The extension to quasi-periodic OL and the  Anderson localization of BEC with effective SOC have been considered in \cite{kit1}. These studies refer mainly to  the continuous case, and the effects of SOC on lattice models (BEC arrays) is practically uninvestigated.

The aim of the present letter is to consider BEC mixtures with SOC in the presence of deep optical lattices. We use Wannier functions to derive a tight-binding model that includes the SOC at the discrete level in the form of a generalized discrete nonlinear Sch\"odinger equation (SO-DNLS) with typical double minima linear dispersion relations, supporting nonlinear  localized two component excitations with chemical potentials inside the forbidden zone of the band structure (gap-solitons), in the nonlinear case. The existence and stability  of gap-solitons (GS) has been investigated as a function of parameters such as the  strength of the interatomic interactions  and
strength of the SO coupling. The band structure and the location of the modes in the band structure is shown to be symmetric with respect to the change of the signs of both inter- and intra- species interactions.  We find that for attractive (resp. repulsive) nonlinear interactions, GS can exist in the semi-infinite gap below (resp. above) the bottom (resp. top) band, as well as in the  gap between the two bands. The time evolution of attractive (resp. repulsive) fundamental GSs, e.g. the one with the lower (resp. higher) chemical potential, are very stable while the ones located in the intraband gap are typically unstable. Quite interestingly, by increasing the strength of the intra-species  interaction $\gamma$, (for simplicity assumed equal for both species)  away from the linear limit, for a fixed and equal sign of  inter-species nonlinearity,  we find three distinctive regions in which GS undergoes spontaneously symmetry breaking. More precisely,  in the range  $0<|\gamma|<|\gamma_1|$ the GS are found to be  asymmetric with respect to the lattice points, in the interval $|\gamma_1| <|\gamma|< |\gamma_2|$ they  display inter-site symmetry (eg symmetry with respect to the middle point between two consecutive lattice sites) and above $|\gamma_2|$ display the on-site symmetry, this behavior being observed both for attractive and repulsive case.  Asymmetric modes appear as novel modes  induced by the SOC.

A possible physical interpretation of this phenomenon is suggested which could be valid also for the corresponding continuous case.

{\it{Tight-Binding Model.\;} }
We consider  BEC with equal contributions of Rashba and Dresselhaus SOC that can be described in the mean field approximation by the following coupled Gross-Pitaevskii equations\cite{sol2}:
\begin{eqnarray}
i \frac{\partial \psi_{j}}{\partial t} & =& (-\frac{\partial^2}{\partial x^2} + V(x)) \psi_{j} - i \alpha \frac{\partial \psi_{3-j}}{\partial x} +  \Omega_j \psi_{j} + \nonumber \\
&& (g_j |\psi_{j}|^2+ g |\psi_{3-j}|^2) \psi_j, \;\;\; j=1,2,
\label{SO-GPE}
\end{eqnarray}
with the linear coefficients $\alpha$, $\Omega\equiv \Omega_1=-\Omega_2$ arising from the  spin orbit interaction while the nonlinear ones,  $g$ and $g_i, i=1,2$,  are related to inter-species and intra-species scattering lengths, respectively. In the following we consider as  trapping potential an optical lattice (OL), e.g. a periodic potential of the form $V(x)=V_0 cos(2x)$ (equal for the two components), and concentrate on the case of  large amplitudes $V_0>>1$ (deep optical lattice) for which it is possible to develop the tight binding approximation\cite{AKKS}. To this regard we expand the two component fields in terms of the Wannier  functions $w(x-n)$ localized around lattice sites $n$ of the underlying uncoupled  linear periodic eigenvalue problem, e.g. with $\alpha=\Omega=0, g=g_i=0 $, in Eq. (\ref{SO-GPE}),
\begin{equation}
\Big[-\frac{\partial^2}{\partial x^2} + V(x)\Big] \varphi_{m,k}=\varepsilon_m(k) \varphi_{m,k}.
\label{periodic-prob}
\end{equation}
Here $\varphi_{m,k}$ and  $\varepsilon_m(k)$ denote, as usual,  Bloch (Floquet) functions and energy band, with $m$ the band index and $k$ the Bloch wavenumber taken in the first Brillouin zone. For a deep optical lattice it is convenient to use the  Wannier basis and expand the fields as
\begin{equation}
\psi_1=\sum_{n,m}u_n(t)w_m(x-n),\; \psi_2=\sum_{n,m}v_n(t)w_m(x-n),
\label{Wexp}
\end{equation}
where $w_{m}(x-n)$ denote  Wannier functions associated to the periodic problem (\ref{periodic-prob}),
and $u_n$ , $v_n$ are time dependent expansion coefficients to be fixed in such a manner that Eq. (\ref{SO-GPE}) are satisfied. Notice that in the expansions (\ref{Wexp}) appear the same Wannier functions since the  underlying periodic problem is the same for both components. Also, we remark that Wannier functions are  orthonormal with respect to both the band index and the lattice site $n$ around which they are centered:
\begin{equation}
\int{w_j(x-n)^* w_{j'}(x-n') dx =\delta_{j,j'} \delta_{n,n'} }
\label{ortonorm}
\end{equation}
In the following we restrict only to one band (the ground state band) and therefore we drop out  the band index $m$. By substituting (\ref{Wexp}) in Eq. (\ref{SO-GPE}) and projecting the
resulting equations along the $w(x-n)$ function, one arrives at the following coupled SO- DNLS system for the coefficients

\begin{equation}
\begin{split}
i\frac{d u_n}{dt}=&\,-\Gamma(u_{n+1}+u_{n-1})+ i\frac{\sigma}{2}(v_{n+1}-v_{n-1})+\Omega u_n+ \\
&\, (\gamma_1 |u_n|^2+ \gamma |v_n|^2) u_n,\\
i \frac{d v_n}{dt}=&\, -\Gamma (v_{n+1}+v_{n-1}) + i\frac{\sigma}{2} (u_{n+1}-u_{n-1}) - \Omega v_n + \\
&\, (\gamma |u_n|^2+ \gamma_2 |v_n|^2) v_n,
\label{SO-DNLS}
\end{split}
\end{equation}
with
\begin{eqnarray}
&&\Gamma \equiv \Gamma_{n,n+1}=\int w(x-n)^*\frac{\partial^2}{\partial x^2}w(x-(n+1),\nonumber \\
&&\gamma_i= g_i \int |w(x-n)|^4 dx,\;\; \gamma= g \int |w(x-n)|^4 dx, \\
&&\sigma\equiv \sigma(n,n+1)= 2 \alpha \int w(x-n)^* \frac{\partial}{\partial x} w(x-(n+1)).\nonumber
\end{eqnarray}
In the derivation of Eq. (\ref{SO-DNLS}) the following relations among coefficients have been used:
$$\sigma(n,n)=0,\; \sigma(n,n-1)=-\sigma(n-1,n)=-\sigma(n,n+1),$$  $$\Gamma(n,n+1)=\Gamma(n,n-1).$$ Moreover, due to the strong localization of the Wannier functions around the lattice sites,  the sums on $n$  have been restricted to onsite and to next neighbor sites  only, for non diagonal and diagonal terms, respectively.

{\it Linear Case.}
Let us first consider the case with all nonlinear coefficients detuned to zero, $\gamma=\gamma_i=0$,
{\small
\begin{eqnarray}
&&i\frac{d u_n}{dt}=-\Gamma(u_{n+1}+u_{n-1})+i\frac{\sigma}{2}(v_{n+1}-v_{n-1})+\Omega u_n, \\
&&i\frac{d v_n}{dt}=-\Gamma(v_{n+1}+v_{n-1})+i\frac{\sigma}{2}(u_{n+1}-u_{n-1})-\Omega v_n.
\label{SO-LSE}
\end{eqnarray}
}
corresponding to the case of a linear chain with spin-orbit coupling. From physical point of view this correspond to a one-dimensional array of BEC in deep optical lattice in the presence of  SO interaction. To find the dispersion relation of this linear chain we consider solutions of the form
\begin{equation}
u_n= A e^{i( k n a - \omega t)},\;\;\;v_n= B e^{i(k n a - \omega t)}
\label{linear-modes}
\end{equation}
with wavenumber $k_n =2 \pi/L n $ varying in the first Brillouin zone $[-\pi/a, \pi/a]$,  with $-N/2\le n \le N/2$,  $L=N a$ the length of the chain, $a$ the lattice constant fixed below without loss of generality to $a=1$. Substituting (\ref{linear-modes}) in Eq. (\ref{SO-LSE}) one obtains a homogeneous system of equations for the coefficients $A, B$, whose compatibility conditions  directly leads to the dispersion relation
\begin{equation}
\omega(k)_\pm = -2 \Gamma \cos(k) \pm \sqrt{\Omega^2 + \sigma ^2 \sin^2(k)}.
\end{equation}
\begin{figure}
\centerline{\includegraphics[scale=0.5]{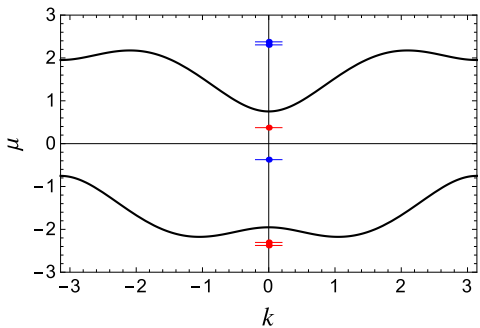}}
\caption{Chemical potential vs $k$ in the reciprocal space for the  DNLS chain with SO coupling for the linear case $\gamma_1=\gamma_2=\gamma=0$ and parameter values $\Gamma =0.3, \Omega=1.352, \sigma=1.5$. The red and blue dots  represent chemical potentials  of localized  modes for the nonlinear cases: $\gamma_1=\gamma_2=-0.65, \gamma_{12}=-1.8$ (attractive case), and   $\gamma_1=\gamma_2=0.65, \gamma_{12}=1.8$ (repulsive case), respectively. Modes are plotted on the $k=0$ line for graphical convenience.}
\label{fig1}
\end{figure}
\begin{figure}
\vskip 1cm
\centerline{\includegraphics[scale=0.385]{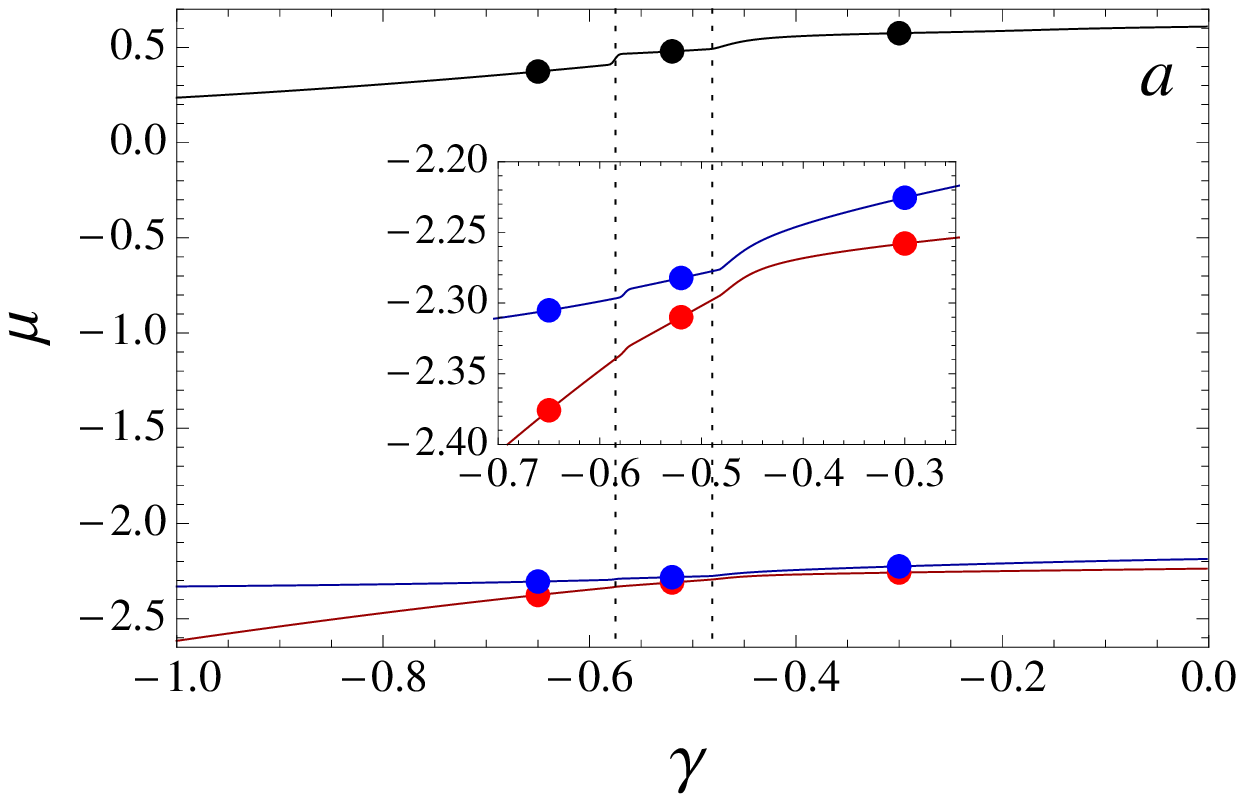}
\hskip -.25cm
\includegraphics[scale=0.365]{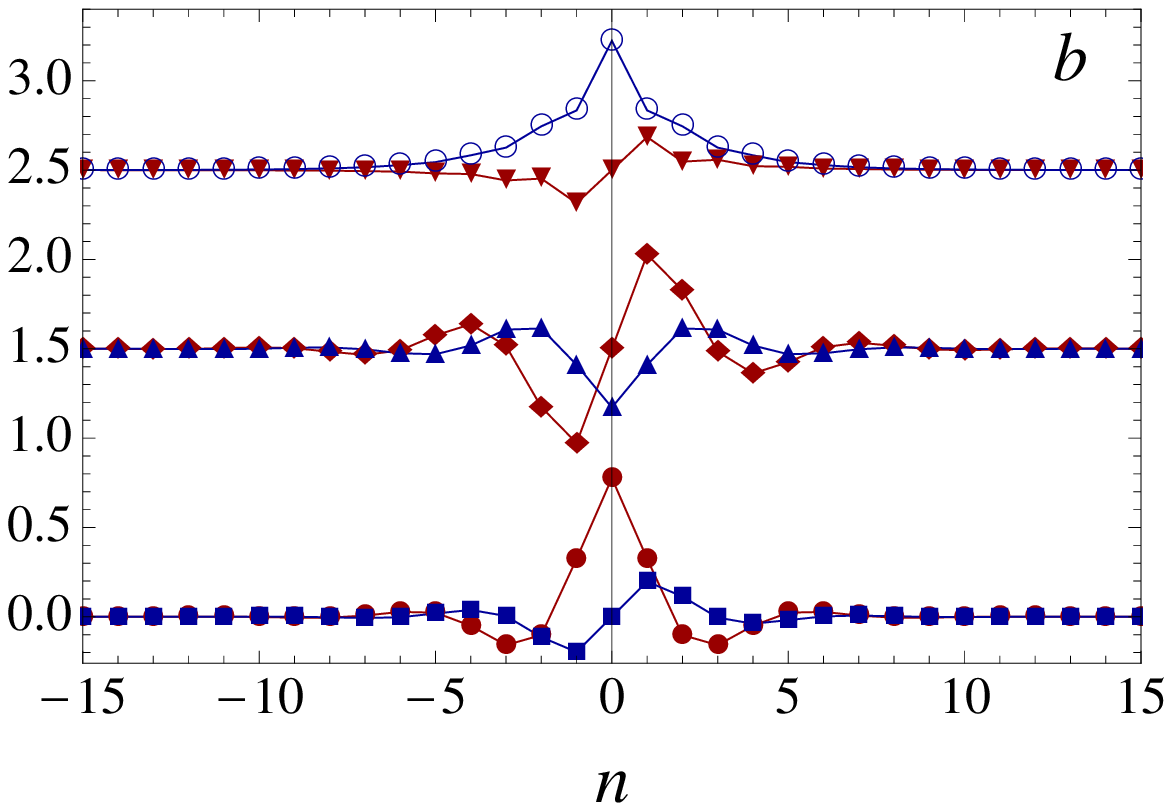}
}
\centerline{\includegraphics[scale=0.365]{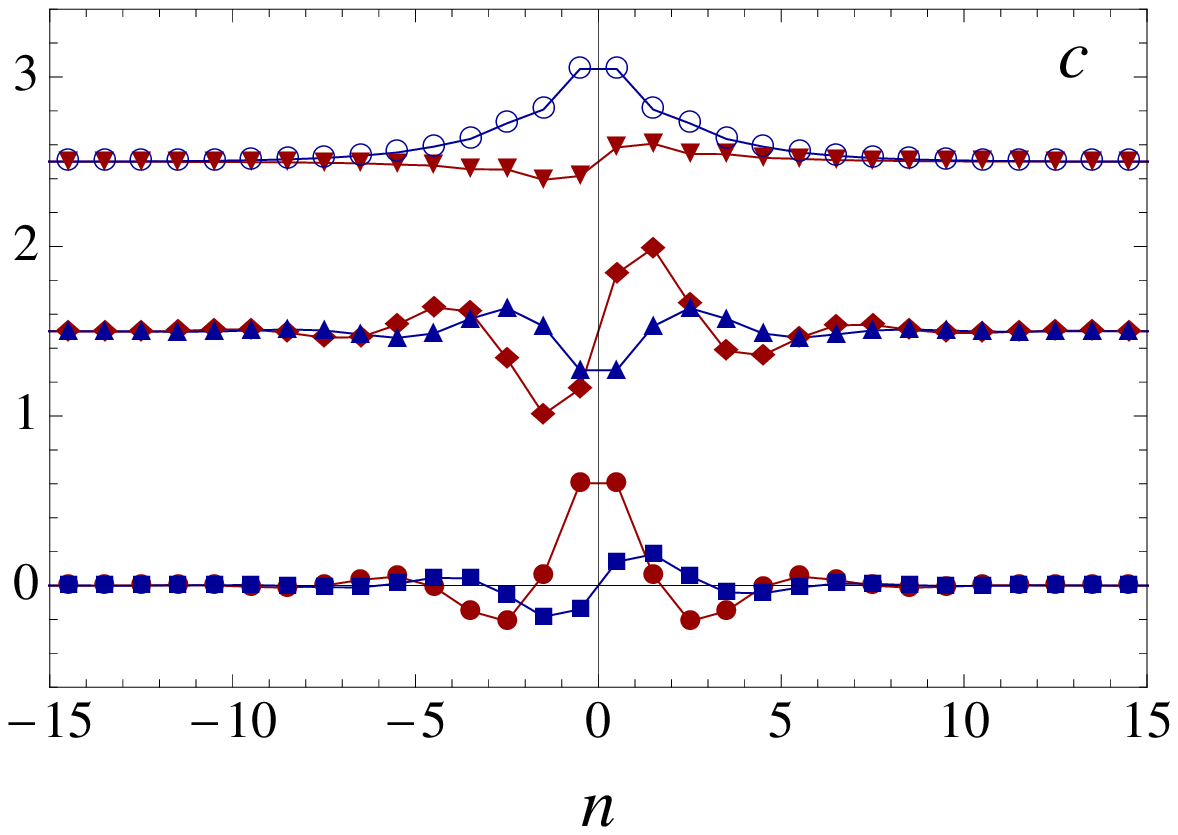}
\hskip -.25cm
\includegraphics[scale=0.365]{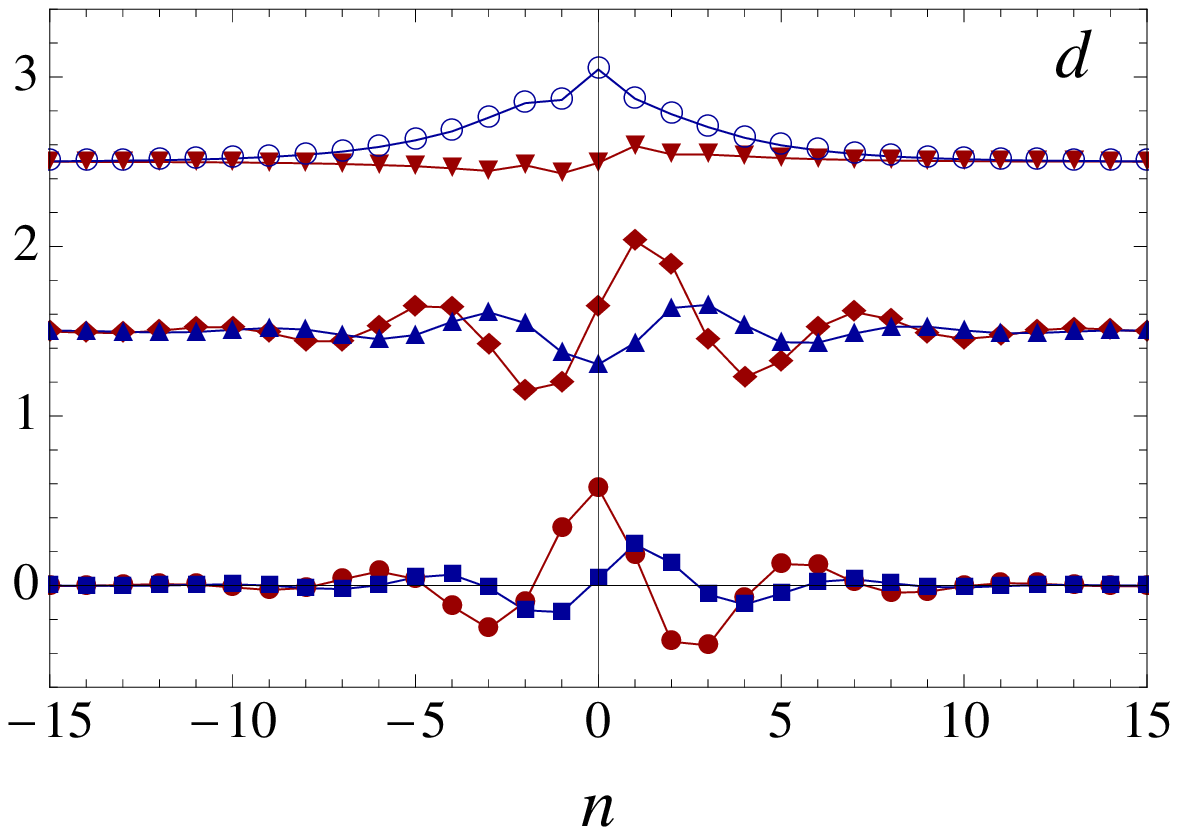}
}
\caption{Panel (a). Existence curves of the localized modes of the DNLS with SO coupling as a function
of the equally attractive intraspecies interaction $\gamma \equiv\gamma_1=\gamma_2$ for parameters $\Gamma=0.3, \Omega=1.352, \sigma=1.5$ and fixed intraspecies interaction $\gamma_{12}=-1.8$. The inset display details of the bottom curves, enlarged along the vertical axis. Panels (b), c and (d) show $u_n$ (blue lines) and $v_n$ (red lines) profiles of gap solitons at $\gamma=-0.65$ (panel (b)), $\gamma=-0.52$ (panel (c)) and  $\gamma=-0.3$ (panel (d)),  corresponding to red points (bottom profiles), blue points (middle profiles) and black points (top profiles) shown in panel (a) (in all cases $v_n$ is purely real while  $u_n$ is purely imaginary). Middle and top  profiles in all panels are shifted upward by $1.5$ and $2.5$, respectively, to avoid overlappings. Vertical dotted lines separate regions of different symmetry type. All plotted quantities are in dimensionless units.}
\label{fig2}
\end{figure}

It is worth to note that the upper and lower branches of the dispersion curve are related as follows
\begin{equation}
\omega_\pm(k)=- \omega_\mp(k+\pi).
\end{equation}
\begin{figure}
\centerline{
\includegraphics[scale=0.25]{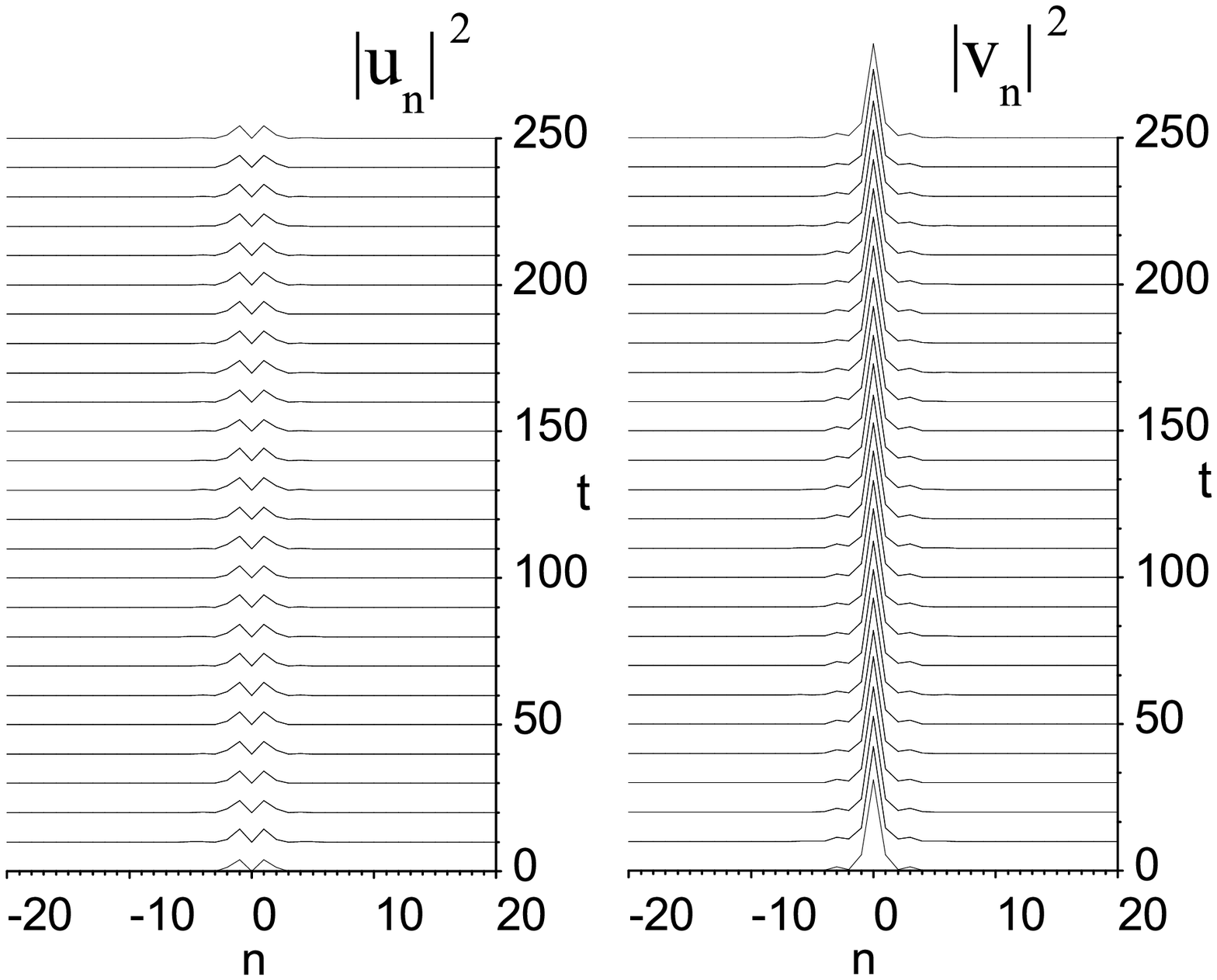}
}
\vskip -2.5cm
\centerline{
\includegraphics[scale=0.25]{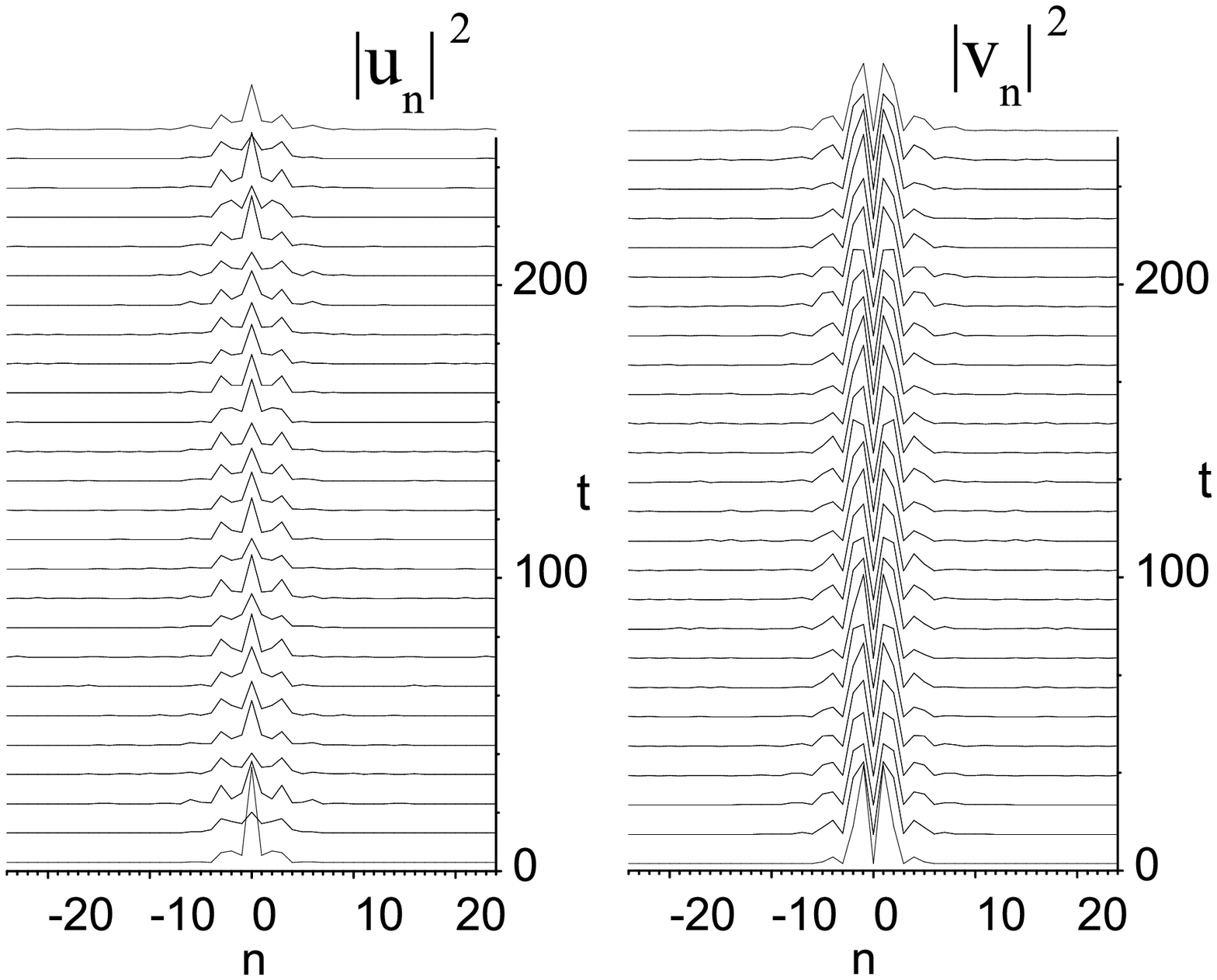}
}
\vskip -2.5cm
\caption{Time evolution of the localized discrete modes shown in panels b of Fig. \ref{fig2} to the parameter value  $\gamma=-0.65$.Top and bottom panels refer to the ground state and to the  first excited state in the lower semi-infinite gap, respectively. Other parameters are fixed as in Fig. \ref{fig2}.}
\label{fig3}
\end{figure}
Also notice the presence of two degenerated minima at $\pm k_{-}$ on the ground state branch and two degenerated maxima at $\pm k_{+}$ on the upper branch, with $k_\pm$ given by
\begin{equation}
k\pm =\pm arcos\Big[ \pm \frac{2 \Gamma}\sigma \sqrt{\frac{\sigma^2+\Omega^2}{\sigma^2+ 4 \Gamma^2}}\Big].
\end{equation}
The dispersion curve in the first Brillouin zone is depicted for typical parameter values in Fig. \ref{fig1}.
One can readily show that the amplitudes of the two components must be related by
\begin{equation}
(\frac{B}A)_\pm= \frac{csc(k)}\sigma (\Omega \mp \sqrt{\Omega^2+\sigma^2 \sin^2(k)}).
\end{equation}
By combining the $k_+$ and  $k_-$ modes one readily obtain stationary stripe solutions for $\omega_+$ (top panels of Fig. \ref{fig2}) and $\omega_-$ (bottom panels of Fig.\ref{fig2}).
\begin{figure}
\centerline{
\includegraphics[scale=0.25]{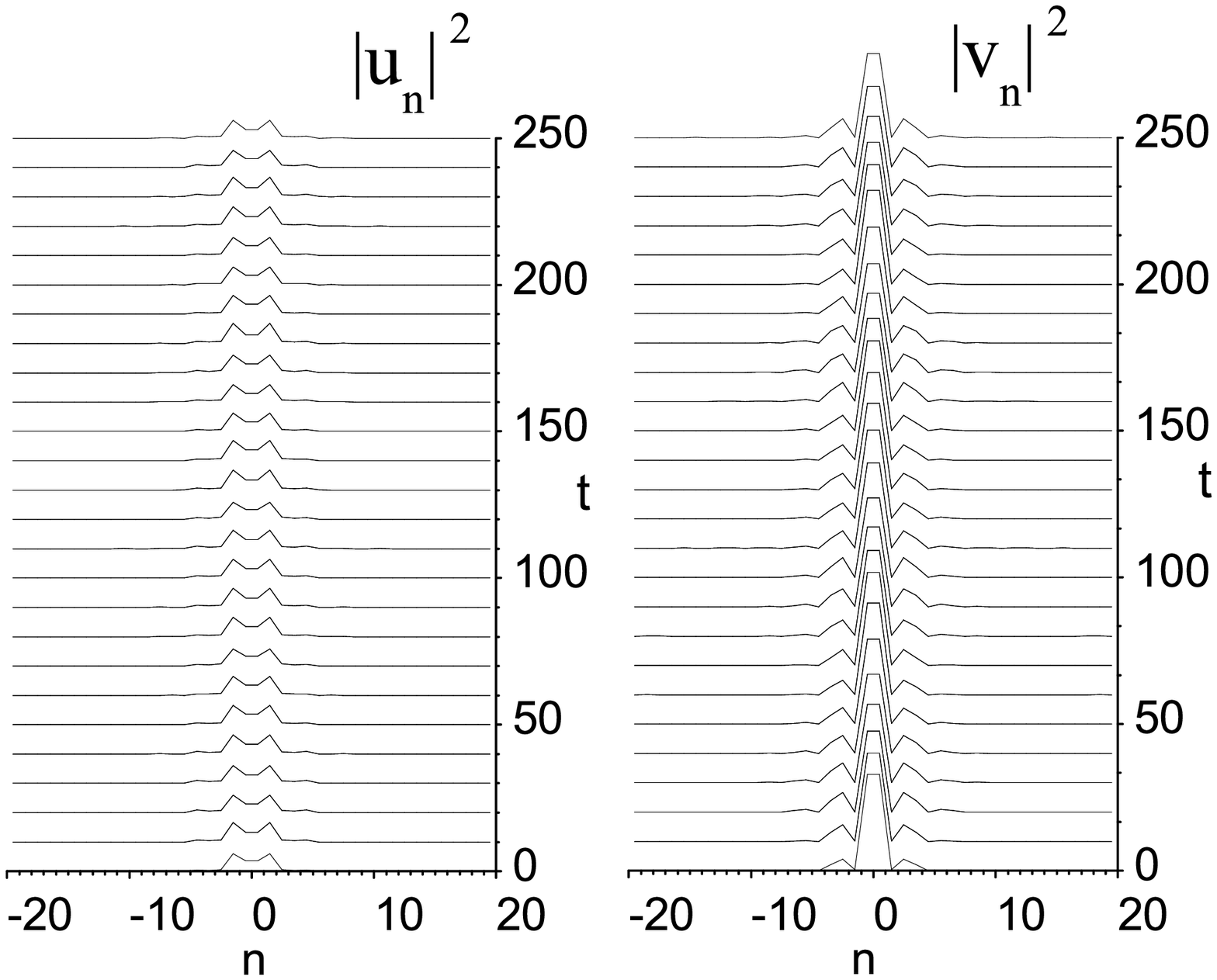}
}
\vskip -2.5cm
\centerline{
\includegraphics[scale=0.25]{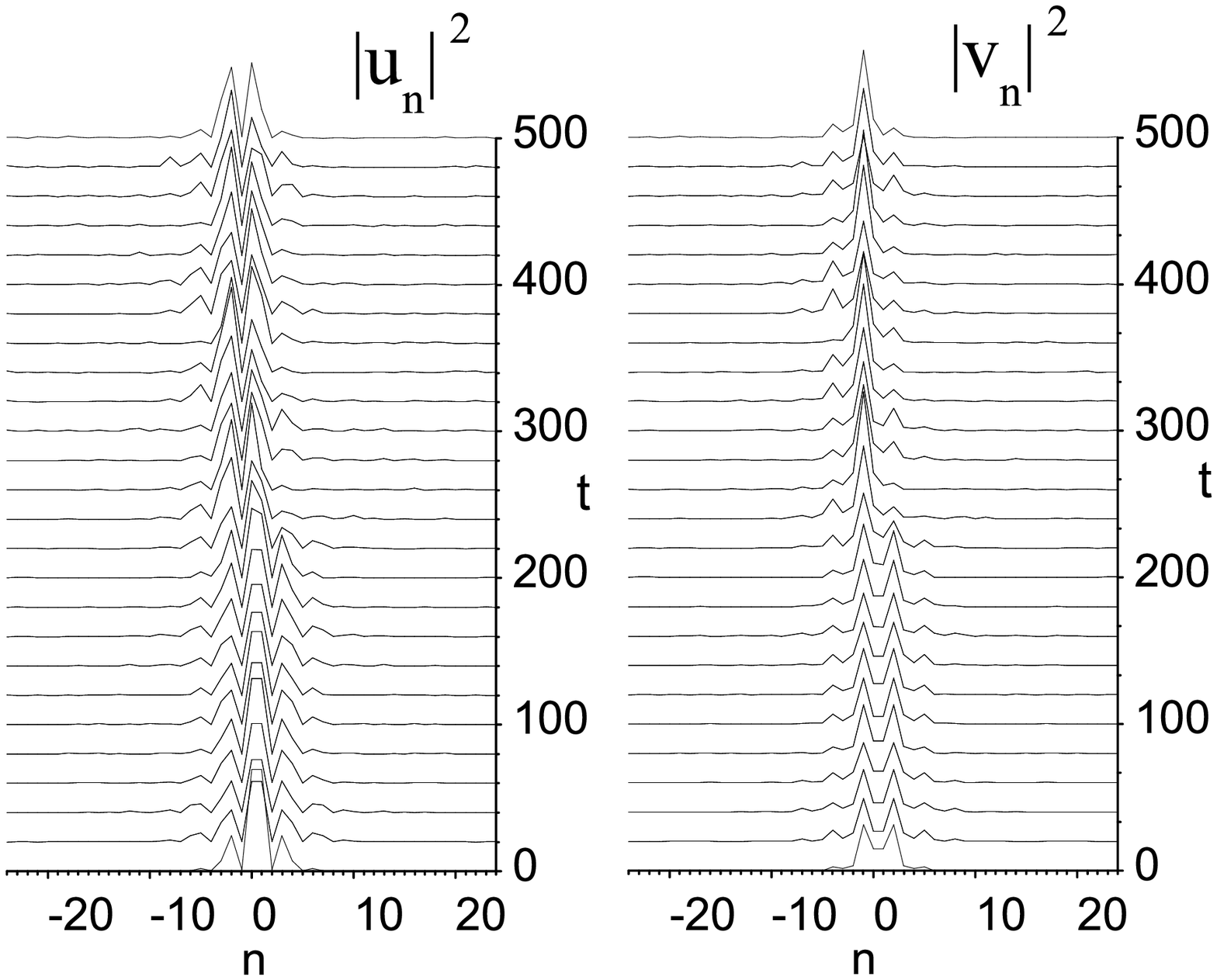}
}
\vskip -2.5cm
\caption{Time evolution of the localized discrete modes shown in the panel c of Fig. \ref{fig2} for the case $\gamma=-0.52$. Top and bottom panels refer to the ground state and to the  first excited state in the lower semi-infinite gap, respectively. Other parameters are fixed as in Fig. \ref{fig2}.}
\label{fig4}
\end{figure}
\begin{figure}
\centerline{
\includegraphics[scale=0.25]{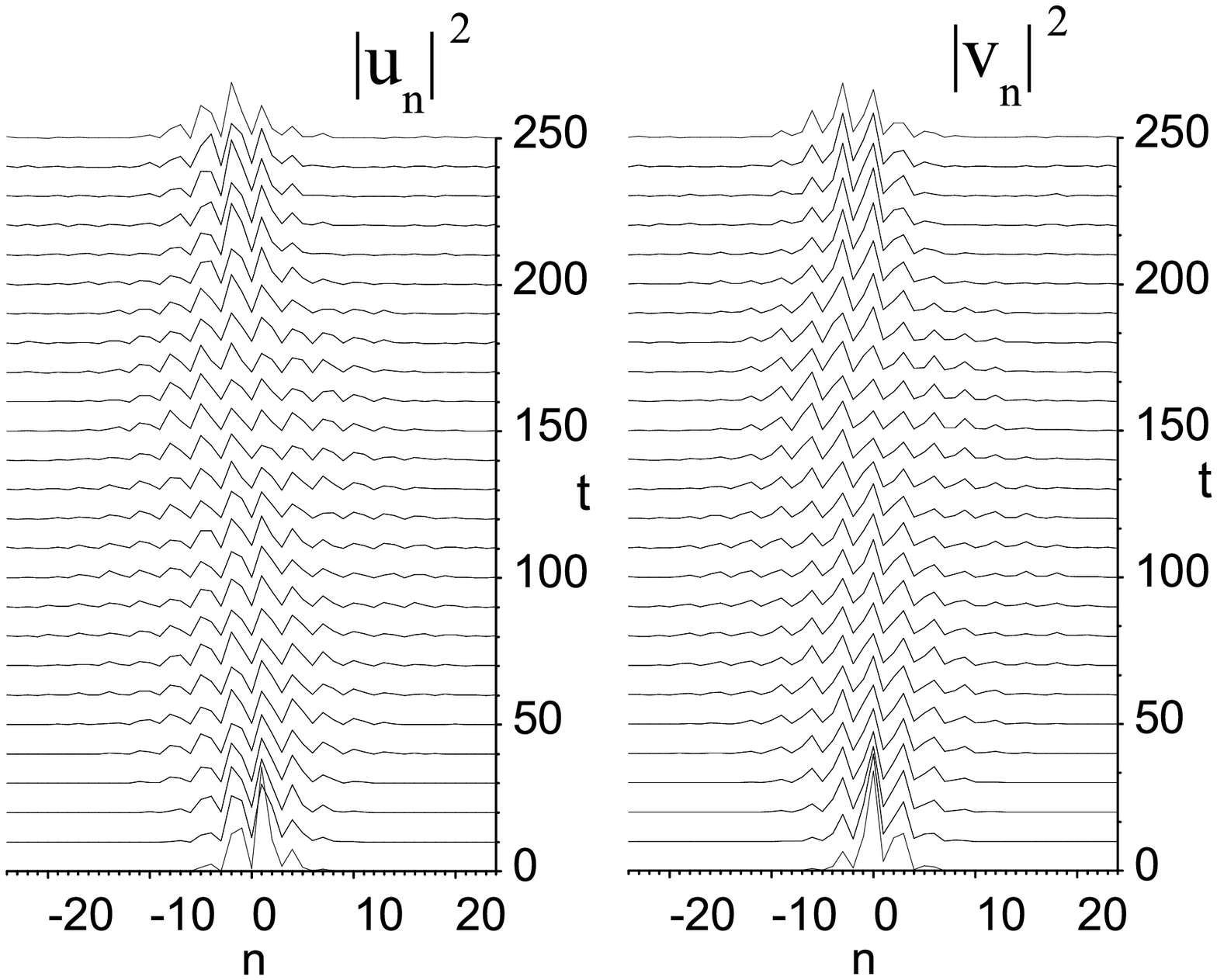}
}
\vskip -2.5cm
\centerline{
\includegraphics[scale=0.25]{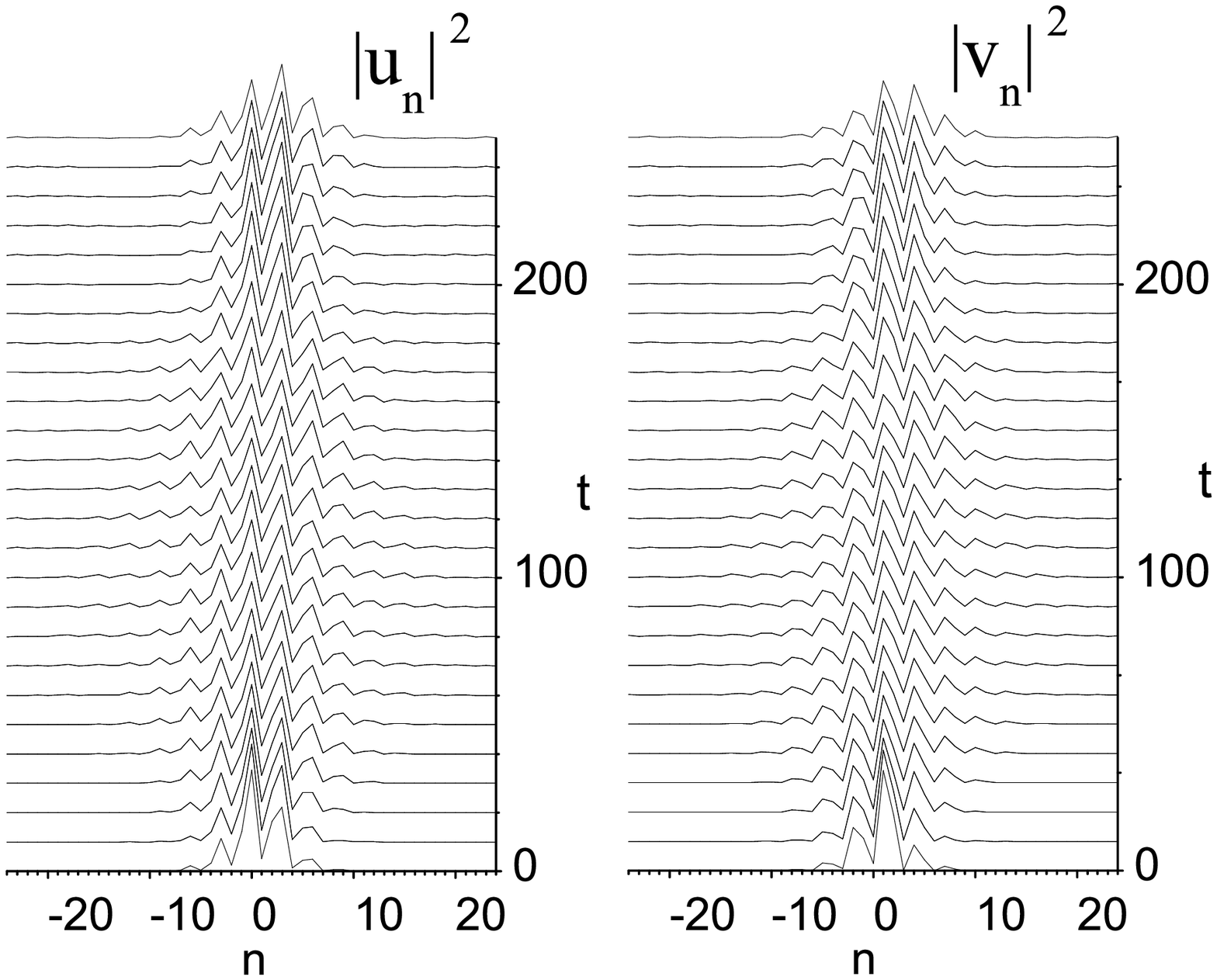}
}
\vskip -2.5cm
\caption{Time evolution of the localized discrete modes shown in the panel d of Fig. \ref{fig2} for the case $\gamma=-0.3$. Top and bottom panels refer to the ground state and to the  first excited state in the lower semi-infinite gap, respectively. Other parameters are fixed as in Fig. \ref{fig2}.}
\label{fig5}
\end{figure}

{{\it Nonlinear Case.\;}}
In the presence of the nonlinearity the existence of localized modes
with chemical potentials in the forbidden zones  of the bandgap structure (discrete gap-solitons) become possible.

In particular, in analogy with gap solitons or intrinsic localized modes of continuous and discrete Gross-Pitaevskii equations in absence of spin orbit coupling, one expects that for attractive (resp. repulsive)  inter and intra-species interatomic interactions, gap-solitons to form with chemical potentials just below  local  minima  (maxima) of the dispersion relation. This is indeed what one finds from a numerical self-consistent diagonalization of  the stationary eigenvalue problem associated to Eq. (\ref{SO-GPE}) (see Fig. (\ref{fig1}) for typical examples).

In the panel a) of Fig. \ref{fig2} are shown  existence curves of gap-solitons of the DNLS with SO coupling for the case $\gamma_{12}<0$ and for  equally attractive intra-species interactions $\gamma_1=\gamma_2 \equiv \gamma <0$ . The lower two branches correspond to the ground localized modes in the lower semi-infinite gap while the top curve refers to a mode inside the interband gap.  In remaining panels of the figure are shown the imaginary and real parts of the component profiles $u_n$ (blue lines)  and $v_n$ (red lines) of the gap solitons  for the different values of $\gamma$ corresponding to black, blue and red points depicted in panel (a).

It is worth to note that there is a phase difference of $e^{i\pi/2}$ between the $u_n$ and $v_n$ components ($v_n$ being real and $u_n$ purely imaginary) as well as different  symmetry properties with respect to the lattice sites for the two components. Also notice that the symmetry properties of the modes change as the intra-species interactions are varied. In general, the following situation
is observed. For a fixed attractive inter-species nonlinearity, we find three distinctive regions in which GS undergoes spontaneously symmetry breaking as the strength of the attractive intra-species  interactions (assumed equal e.g. $\gamma_i\equiv\gamma$, for simplicity)  away from the $\gamma=0$ limit.   More precisely,  in the range  $-0.35<\gamma<0$ the GS are found to be  asymmetric with respect to the lattice points, in the interval $-0.6 <|\gamma|<- 0.35$ they  display a symmetry with respect to the middle point between two consecutive lattice sites (inter-site symmetry) and for $\gamma<0.6$ they display the  on-site symmetry. Notice that the borders of these regions correspond to the appearance of small kinks in the existence curves $\mu$ versus $\gamma$ and have been  evidenced by dashed vertical lines (see inset in panel a) of  and Fig. \ref{fig2}). Also notice that  the kinks at the left border of the middle region separating the modes with on-site symmetry (see panel (b)) from the ones with intra-site symmetry (see panel (c)), both of symmetric or anti-symmetric type, are sharper than the ones on the right border where modes becoming asymmetric (see panel (c)). This also correlates with the fact that in general one would expect that a lost of symmetry being a smoother process with respect to a spontaneously change of a symmetry.
\begin{figure}
\centerline{
\includegraphics[scale=0.25]{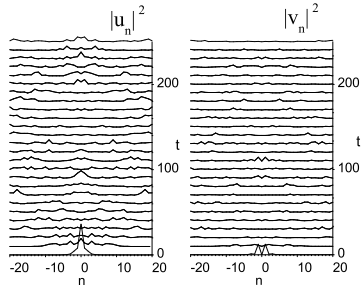}
}
\vskip -2.5cm
\caption{Time evolution of the the on-site discrete gap soliton inside the  inter-band gap shown in the panel b of Fig. \ref{fig2} (top profiles) for the case $\gamma=-0.65$.  Other parameters are fixed as in Fig. \ref{fig2}.}
\label{fig7}
\end{figure}

The dynamical properties of the nonlinear modes depicted in Fig. \ref{fig2}(c)-(d)) have been investigated by direct numerical integrations of the DNLS-SO,  by taking them as initial conditions with a small noise component  imposed to check their stability under time evolution. This is shown in Figs. \ref{fig3}-\ref{fig5} for the two lowest nodes in the semi-infinite gap. As one can see, for the chosen parameters the on-site symmetric modes are both very stable (see Fig. \ref{fig3}) while for inter-site symmetric modes stability is achieved only for the ground state, the first excited state being metastable. Notice, in the last case, that the mode  decays into an on-site- symmetric mode plus background radiation, similarly to the inter-site-on-site transition observed for discrete breathers of coupled DNLS in absence of SO coupling. Quite surprisingly, asymmetric modes of the region $\gamma<-0.3$ also appear to be stable (or long lived) under time evolution, as one can see form Fig. \ref{fig5}. The existence of such modes appears to be typical of the SO coupling, since they would not be possible just in ordinary coupled DNLS. The stability properties of localized  modes in the semi-infinite gap, however, appear to be critical. In particular, for the parameter ranges we have explored and independently from their symmetry types, none of them was found to be stable  (see Fig.\ref{fig7} for a typical example).
\begin{figure}
\vskip 1cm
\centerline{\includegraphics[scale=0.385]{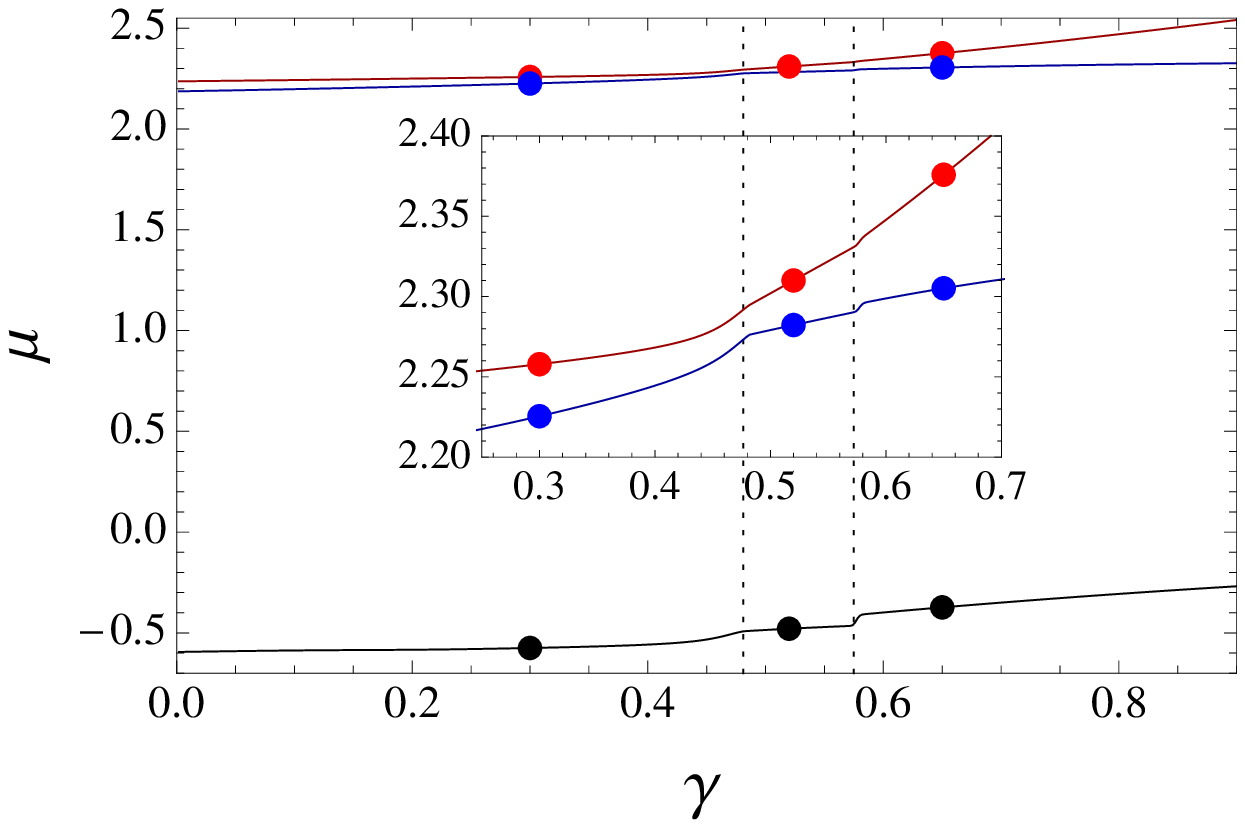}
\hskip -.25cm
\includegraphics[scale=0.365]{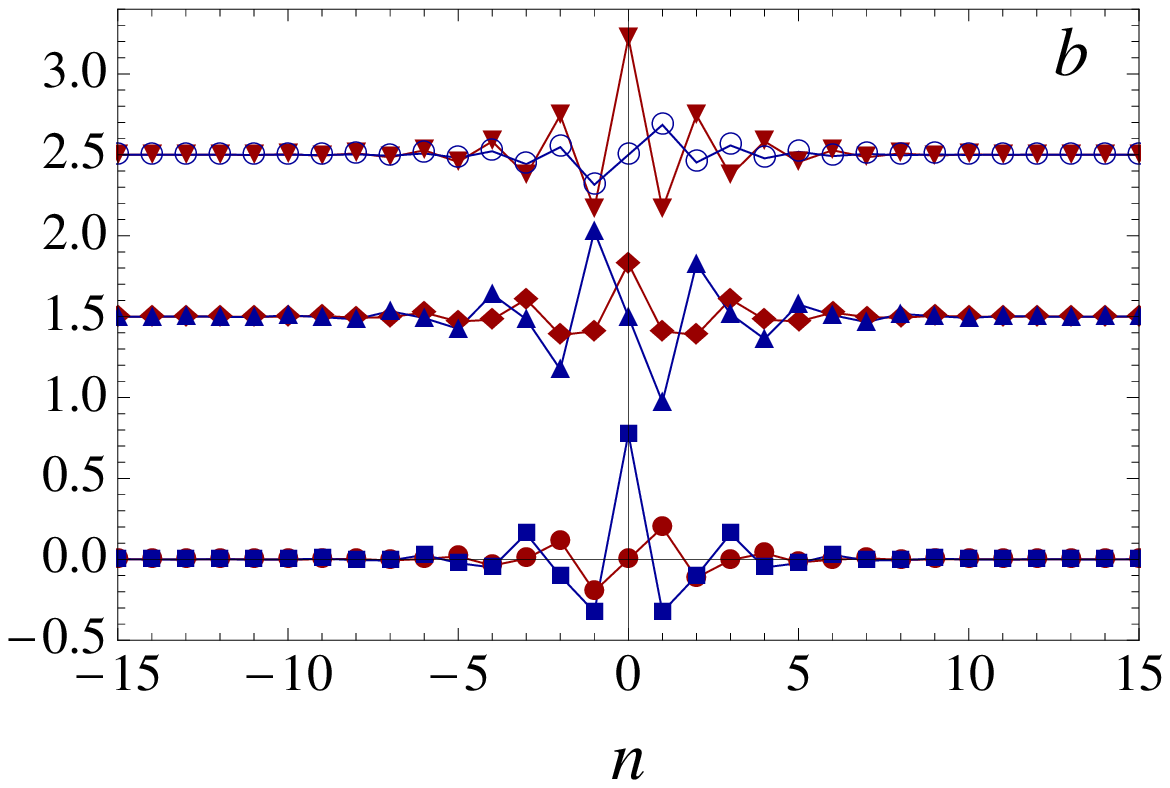}
}
\centerline{\includegraphics[scale=0.365]{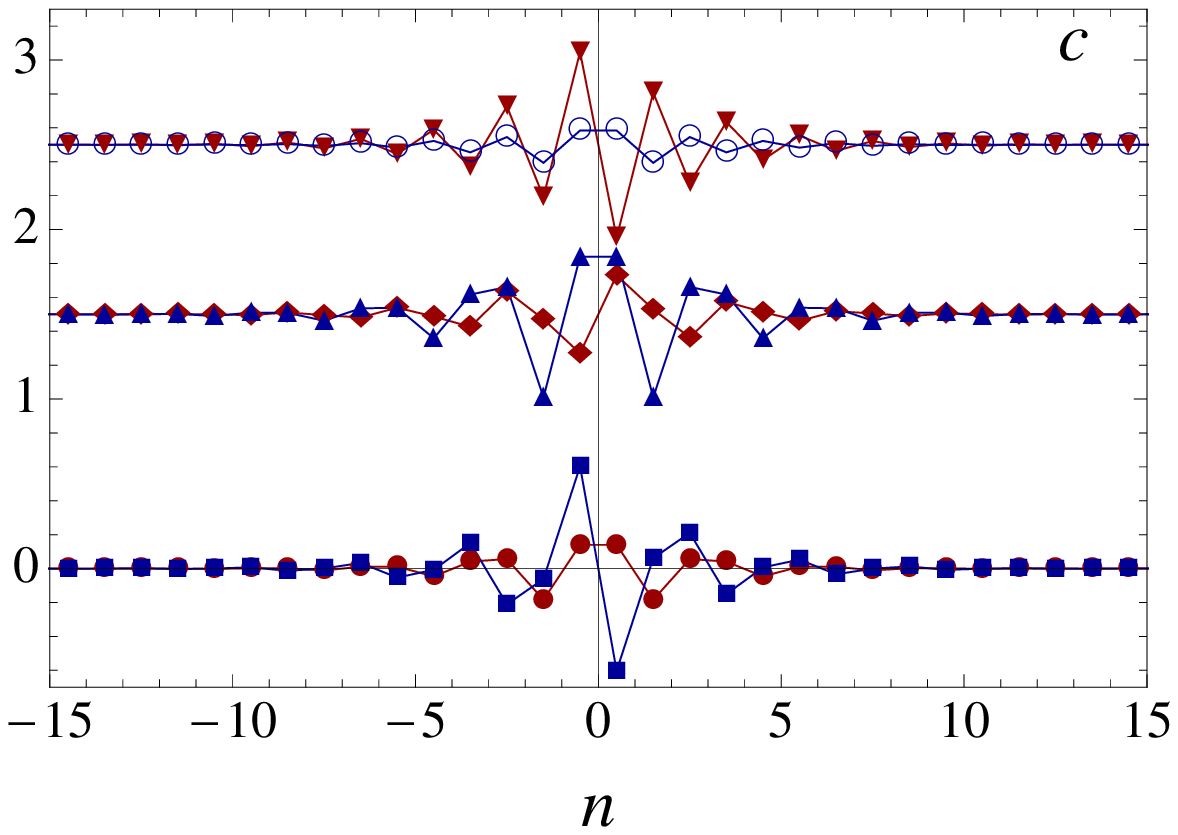}
\hskip -.25cm
\includegraphics[scale=0.365]{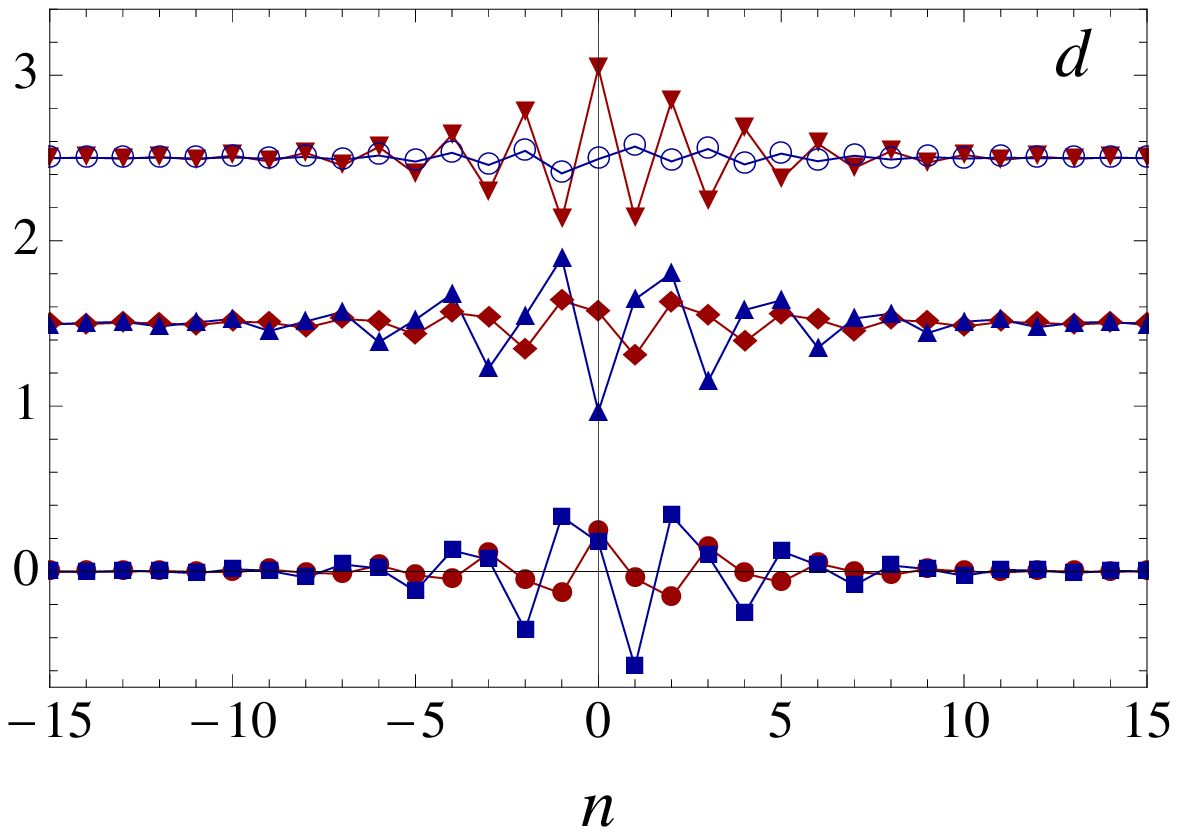}
}
\caption{Panel (a). Existence curves of the localized modes of the DNLS with SO coupling as a function
of the equally repulsive intraspecies interaction $\gamma \equiv\gamma_1=\gamma_2$ for parameters $\gamma=0.3, \Omega=1.352, \sigma=1.5$ and fixed intraspecies interaction $\gamma_{12}=1.8$. The inset display details of the top curves, enlarged along the vertical axis.  Panels (b), (c) and (d) show $u_n$ (blue lines) and $v_n$ (red lines) profiles at $\gamma=0.65$ (panel (b)), $\gamma=0.52$ (panel (c)) and  $\gamma=0.3$ (panel (d)), corresponding to black points (bottom profiles), blue points (middle profiles) and red points (top profiles)  in panel (a), respectively (in all cases $v_n$ is purely real while  $u_n$ is purely imaginary). Middle and top profiles in the b,c,d, panels are shifted upward by $1.5$ and $2.5$, respectively, to avoid overlapping. Vertical dotted lines separate regions of different symmetry type. All plotted quantities are in dimensionless units.}
\label{fig8}
\end{figure}

Similar results are found for the case of all repulsive interactions. By reversing signs of all the interactions  chemical potentials of the discrete gap solitons also change their signs (see Fig. \ref{fig1}). Shape and symmetries of the modes, however,  are different in the two cases as one can see by comparing Figs. \ref{fig8} with \ref{fig2}.  Also notice the symmetry of corresponding existence curves in the $\mu-\gamma$ plane, with appearance of kinks at the change of symmetry points.  The stability  properties under time evolution, look also similar to the all attractive case. In particular, modes inside the inter-band gap are also found to be unstable, while the ones with highest chemical potentials inside the upper semi-infinite gap are typically stable (not shown for brevity).

In closing this Letter we provide parameter estimates for possible experimental implementations of the above results. For this one can adopt an  experimental  setting similar  to the one  in \cite{LJS}. In particular, the SOC can be realized in $^{87}$Rb  using two counter-propagating Raman lasers  with $\lambda_R = 804$nm, spin-orbit coupling $\alpha = \hbar k_R/m$, and $\hbar\Omega = 2E_R$  (here $E_R = \hbar^2 k_R^2/2m$ is the recoil energy). The optical lattice can be generated by means of  two additional beams with $\lambda_L=1540$nm\cite{exp1}. Typical values of the OL potential strength $V_0$ suitable for the tight-binding case we considered can be $V_0/E_R > 10$\cite{AKKS,MO}. In the case of all attractive interspecies interaction nonlinear modes below the bands should appear for wide range of number of atoms, being typically stable. Change of the interspecies scattering length to observe  the change of symmetry of the  discussed modes can be achieve by means of the Feshbach resonance technique.

In conclusion we have derived a tight-binding model for BEC with SOC in deep optical lattices  and demonstrated for this  model the existence and stability of different types of discrete solitons.
We showed that nonlinear  modes can change symmetry from on-site symmetric  to inter-site symmetric and to fully asymmetric as the interspecies interaction is varied. Asymmetric modes  appear to be intrinsic novel excitations of the BEC with SOC.
The critical values at which  the symmetry changes are found to correspond to kink-like profiles in the chemical potential  existence curves that suggest   the occurrence of phase transitions. The possibility to observe these phenomena  in real experiments was suggested.
\vskip 0.25cm
{{\it Acknowledgements}}
 M. S. acknowledges partial support from the Ministero dell'Istruzione, dell'Universit\'a e della Ricerca (MIUR) through a PRIN (Programmi di
Ricerca Scientifica di Rilevante Interesse Nazionale) 2010-2011 initiative.

\end{document}